# Anomalous four-fermion processes in electron-positron collisions [*]

F.A. Berends[†] and A.I. van Sighem[‡]

Instituut-Lorentz, University of Leiden,

P.O.B. 9506, 2300 RA Leiden, The Netherlands

June 22, 1995

**Abstract**

This paper studies the electroweak production of all possible four-fermion states in $e^+e^-$ collisions with non-standard triple gauge boson couplings. All $CP$ conserving couplings are considered. It is an extension of the methods and strategy, which were recently used for the Standard Model electroweak production of four-fermion final states. Since the fermions are taken to be massless the matrix elements can be evaluated efficiently, but certain phase space cuts have to be imposed to avoid singularities. Experimental cuts are of a similar nature. With the help of the constructed event generator a number of illustrative results is obtained for $W$-pair production. These show on one hand the distortions of the Standard Model angular distributions caused by either off-shell effects or initial state radiation. On the other hand, also the modifications of distributions due to anomalous couplings are presented, considering either signal diagrams or all diagrams.

---

[*]This research has been partly supported by EU under contract number CHRX-CT-92-0004.

[†]email address: berends@rulgm0.LeidenUniv.nl

[‡]email address: h88@nikhef.nl

# 1 Introduction

Recently, the electroweak four-fermion production processes relevant for LEP2 and beyond have been studied in a number of ways. One of the objectives is to obtain a description of $W$-pair production better than an on-shell treatment with $W$-decay products attached to it. Thus all recent papers contain finite width effects. Some papers only include the three diagrams leading to $W$-pair production, others include all diagrams giving a specific four-fermion final state. Most of them include some form of initial state QED radiative corrections. There are semi-analytical methods [1, 2] and Monte Carlo approaches [3]–[10]. The former can only give distributions in the virtualities of the $W$'s, but no fermion distributions. The latter can produce any wanted distribution.

Among the various Monte Carlo treatments we mention in particular the program EXCALIBUR, since it aims both at completeness and speed. All diagrams for any four-fermion final state are included and a relatively fast calculation is achieved by assuming massless fermions and by using a multichannel approach to generate the phase space. The details are given in [9], whereas the treatment of initial state radiation (ISR) can be found in [10].

One of the objectives of LEP2 and future electron-positron colliders is a test of non-abelian triple gauge boson couplings. A way to quantify deviations from the Standard Model (SM) Yang-Mills couplings is to set experimental limits on anomalous couplings. Many discussions of the latter can be found in the literature, see e.g. [11, 12, 13]. Theoretical arguments, which reduce the a-priori large number of non-standard couplings are discussed in [13]. In order to investigate the experimental possibilities to measure limits on anomalous couplings one ideally needs samples of anomalous events, made by an event generator and one requires a fitting program containing the anomalous matrix element. The fitting program can then establish whether the input anomalous couplings can really be extracted from the generated anomalous data.

Up to now such studies were made with tools, which have certain limitations. Usually data are generated for $W$-pair production containing three diagrams with $W$-decay attached to it in zero width approximation. The fitting programs use the same approximation. Examples of such investigations can be found in [13, 14]. Very recently a Monte Carlo program with anomalous couplings with a finite $W$-width became available [15]. It covers the semi-leptonic final states.

In view of the advantages of the EXCALIBUR program it is natural to use its structure and strategy as a basis for an anomalous four-fermion generator.

Thus it is the aim of the present paper to describe the necessary additions and changes to the approach of [9, 10] to obtain a four-fermion generator with anomalous couplings.

This generator has the following characteristics. Any massless four-fermion final state can be produced with $CP$ conserving anomalous couplings. All abelian and non-abelian diagrams contributing to the final state are taken into account. It is also possible to restrict oneself to the "signal" diagrams of a process, e.g. for four-fermion production through $W$-pairs one takes only three diagrams. Finally, ISR can be switched on or off.

The new anomalous matrix elements will be discussed in some detail, since they are the key ingredient of the anomalous EXCALIBUR program and since they would also be required for a fitting program.

A number of numerical results for four-fermion production will be presented. On one hand they serve as illustration how the $W$ finite width or ISR can modify SM angular distributions. On the other hand they show how non-standard distributions behave. The very relevant physics application is the one mentioned above: generating anomalous data and studying the extraction of anomalous couplings with a fitting program. It is expected that in the future this question will be addressed.

The actual outline of the paper is as follows. In section 2 the anomalous couplings are described. The next section discusses those four-fermion final states, which are sensitive to anomalous couplings and gives the required matrix elements. Some illustrative examples of anomalous effects in distributions are shown in section 4, whereas section 5 contains conclusions.

## 2 Anomalous couplings

In this section those non-standard couplings are defined, which will be considered for the generation of anomalous four-fermion final states. When one uses only Lorentz invariance as condition there exist 14 couplings, which lead to deviations from the SM triple gauge boson couplings. Some of them can be immediately discarded since they would either modify the strength of the electromagnetic interaction or introduce $C$ or $CP$ violation in it. At this point there still are 9 parameters left, three of which lead to $CP$ violation through the $ZWW$ interaction. Also these will be omitted. We are then left with a Lagrangian of the form:

$$\begin{aligned} \mathcal{L} &= \mathcal{L}_1(\mathcal{C}\text{- and } \mathcal{P}\text{-conserving}) + \\ &\quad \mathcal{L}_2(\mathcal{CP}\text{-conserving}, \mathcal{C}\text{- and } \mathcal{P}\text{-violating}). \end{aligned} \qquad (1)$$

Here $\mathcal{L}_1$ takes the form

$$\mathcal{L}_1 = -ie\,[A_\mu(W^{-\mu\nu}W^+{}_\nu - W^{+\mu\nu}W^-{}_\nu) + F_{\mu\nu}W^{+\mu}W^{-\nu}]$$

$$+ie\cot\theta_w\,[Z_\mu(W^{-\mu\nu}W^+{}_\nu - W^{+\mu\nu}W^-{}_\nu) + Z_{\mu\nu}W^{+\mu}W^{-\nu}]$$

$$-iex_\gamma F_{\mu\nu}W^{+\mu}W^{-\nu} + iex_Z Z_{\mu\nu}W^{+\mu}W^{-\nu}$$

$$+ie\delta_Z\,[Z_\mu(W^{-\mu\nu}W^+{}_\nu - W^{+\mu\nu}W^-{}_\nu) + Z_{\mu\nu}W^{+\mu}W^{-\nu}] \quad (2)$$

$$+ie\frac{y_\gamma}{M_W^2}F^{\nu\lambda}W^-_{\lambda\mu}W^{+\mu}{}_\nu$$

$$-ie\frac{y_Z}{M_W^2}Z^{\nu\lambda}W^-_{\lambda\mu}W^{+\mu}{}_\nu$$

whereas the second part reads

$$\mathcal{L}_2 = -\frac{ez_Z}{M_W^2}\,\partial_\alpha \hat{Z}_{\rho\sigma}\,(\partial^\rho W^{-\sigma}W^{+\alpha} - \partial^\rho W^{-\alpha}W^{+\sigma} + \partial^\rho W^{+\sigma}W^{-\alpha} - \partial^\rho W^{+\alpha}W^{-\sigma}) \quad (3)$$

where $\hat{Z}_{\mu\nu}$ is the dual field tensor

$$\hat{Z}_{\rho\sigma} = \frac{1}{2}\epsilon_{\rho\sigma\alpha\beta}Z^{\alpha\beta} \quad (4)$$

with $\epsilon_{0123} = -1$. Taking the anomalous couplings vanishing leaves us with the first and second lines in $\mathcal{L}_1$, i.e. the SM Lagrangian. Although this general form will be considered, there are theoretical symmetry arguments to reduce the number of independent couplings [13]. In practical fits this reduction will be necessary. For completeness we list the Feynman rule for the $ZWW$ vertex when all particles are considered to be outgoing:

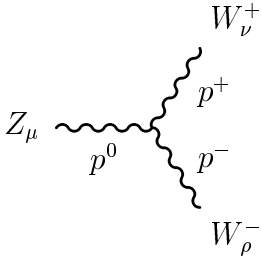

$$\begin{aligned}
= \ &ie\left(\cot\theta_w + \delta_Z\right)\left[g_{\rho\nu}(p^- - p^+)_\mu + g_{\mu\nu}(p^+ - p^0)_\rho + g_{\mu\rho}(p^0 - p^-)_\nu\right] \\
&+ iex_Z\left[g_{\mu\rho}p^0_\nu - g_{\mu\nu}p^0_\rho\right] \\
&+ \frac{iey_Z}{M_W^2}\left[(p^+_\rho p^0_\nu p^-_\mu - p^+_\mu p^0_\rho p^-_\nu) + (p^0 \cdot p^+)(g_{\mu\rho}p^-_\nu - g_{\rho\nu}p^-_\mu)\right. \\
&\qquad\qquad \left. + (p^0 \cdot p^-)(g_{\rho\nu}p^+_\mu - g_{\mu\nu}p^+_\rho) + (p^- \cdot p^+)(g_{\mu\nu}p^0_\rho - g_{\mu\rho}p^0_\nu)\right] \\
&- \frac{ez_Z}{M_W^2}\left[\epsilon_{\mu\rho\kappa\sigma}p^0_\nu - \epsilon_{\mu\nu\kappa\sigma}p^0_\rho\right]p^{0\kappa}(p^+ - p^-)^\sigma .
\end{aligned}$$

For the $\gamma WW$ vertex one has to replace $\cot\theta_w + \delta_Z$ by $-1$, $x_z$ and $y_z$ by $-x_\gamma$ and $-y_\gamma$. The coupling $z_\gamma$ is zero.

With these vertices the matrix elements for four-fermion production will be evaluated. It should be noted that the form chosen in the interaction corresponds to that of [13]. In the $Z$ couplings the signs look different from [13]. This is however compensated by the vectorboson-fermion couplings, which differ between the two papers. For the SM we use here for the photon-electron vertex $ie\gamma^\mu$ and for the $Z$-vertex $ie\gamma^\mu(v - a\gamma_5)$ with $a = -(4\sin\theta_w\cos\theta_w)^{-1}$ and $v = a(1 - 4\sin^2\theta_w)$. In [13] the latter is the same but the photon vertex has opposite sign.

## 3 The matrix elements

In the literature [16] studies have been made of the effect of non-standard triple gauge boson couplings on the following gauge boson production processes

(1) $e^+\ e^- \ \rightarrow \ W^+\ W^-$ ,

(2) $e^+\ e^- \ \rightarrow \ W\ e\ \nu_e$ ,

(3) $e^+\ e^- \ \rightarrow \ Z\ \nu_e\ \bar{\nu}_e$ .

They are described by 3, 9 and 7 diagrams respectively of which 2, 2 and 1 diagrams containing a triple gauge boson vertex. In practice these processes lead to four-fermion final states. For a specific four-fermion final state not only the "signal" diagrams of the above reactions contribute but also "background" diagrams, of which some contain also triple gauge boson vertices. Thus the anomalous couplings can contribute to the background diagrams as well. Tables 1–3 list the leptonic, semileptonic and hadronic final states which can originate from one of the above signals. Moreover, the number of abelian diagrams ($N_a$) and of non-abelian diagrams ($N_n$) is given. For the actual calculation of the matrix element we proceed as in [9]. We first repeat the SM calculation and then extend it to non-standard couplings. Although many diagrams can contribute to a specific final state there are only two topological structures (generic diagrams), given in fig. 1. In these diagrams all particles are considered to be outgoing. The actual Feynman diagrams will be obtained by crossing those electron and positron lines which were assigned to become the colliding $e^+e^-$ pair. In the Abelian diagrams the charges of the fermions determine the character of the two exchanged bosons, which may be $W^+$, $W^-$, $Z$ or $\gamma$. In the non-abelian diagrams, two of the vector bosons are fixed to be $W^+$ and $W^-$, and the third one can be $Z$ or $\gamma$. In this way we avoid double-counting of diagrams. In principle the particles and antiparticles can each be assigned in six ways to the external lines. For the non-abelian diagrams we get at most 8 diagrams and due to the specific final states considered we get at most 48 abelian diagrams. Fixing a specific four-fermion final state all possible assignments are tried. Only those which are allowed for by the couplings survive in first instance. Since also successively all helicity combinations are tried certain diagrams do not contribute as can be seen from the numerator of the generic abelian diagram:

$$\begin{aligned}
A(\lambda, \rho, \sigma; p_1, p_2, p_3, p_4, p_5, p_6) &= \\
&= \bar{u}_\lambda(p_1) \gamma^\mu u_\lambda(p_2) \\
&\times \bar{u}_\rho(p_3) \gamma_\mu (\not{p}_1 + \not{p}_2 + \not{p}_3) \gamma_\nu u_\rho(p_4) \\
&\times \bar{u}_\sigma(p_5) \gamma^\nu u_\sigma(p_6) \ .
\end{aligned} \quad (5)$$

Here we have disregarded the particle/antiparticle distinction since it is already implied by the assignment of the external momenta. The helicity labels $\lambda, \rho, \sigma = \pm 1$ determine the helicity of both external legs on a given fermion line. Using the Weyl-van der Waerden formalism for helicity amplitudes [17] (or, equivalently, the Dirac formalism of [18]), the expression $A$ can easily be calculated. For instance, for $\lambda = \rho = \sigma = 1$ one finds

$$A(+,+,+;1,2,3,4,5,6) = 4\langle 31 \rangle^* \langle 46 \rangle \left[ \langle 51 \rangle^* \langle 21 \rangle + \langle 53 \rangle^* \langle 23 \rangle \right] \quad (6)$$

where the spinorial product is given, in terms of the momenta components, by

$$\langle kj \rangle = \left( p_j^1 + i p_j^2 \right) \left[ \frac{p_k^0 - p_k^3}{p_j^0 - p_j^3} \right]^{1/2} \ - \ (k \leftrightarrow j) \ . \quad (7)$$

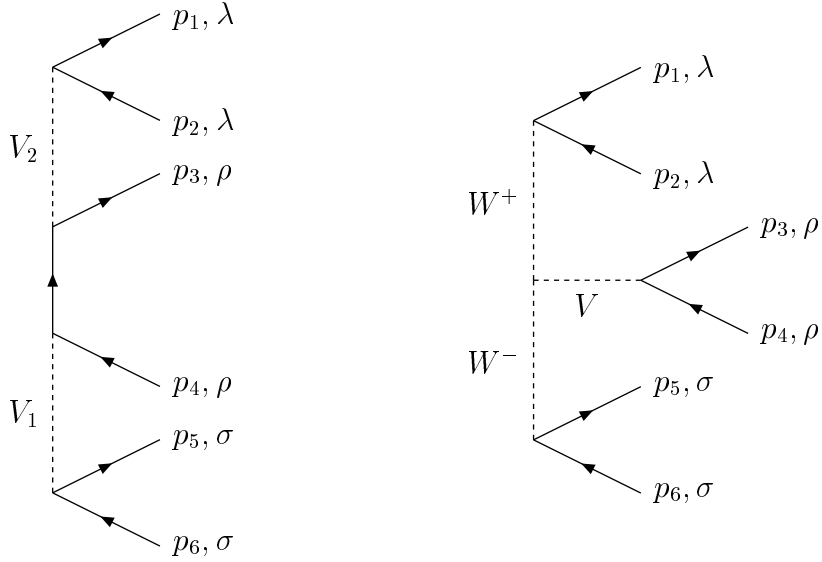

Figure 1: generic diagrams for four-fermion production. The fermion momenta and helicities, and the bosons are indicated. The bosons $V_{1,2}$ can be either $Z$, $W^{\pm}$, or $\gamma$; $V$ can be either $Z$ or $\gamma$.

We denote the expression of equation 6 by $A_0(1,2,3,4,5,6)$. All helicity combinations can be expressed in terms of $A_0$, as follows:

$$\begin{aligned}
A(+++) &= A_0(1,2,3,4,5,6) & A(---) &= A_0(1,2,3,4,5,6)^* \\
A(-++) &= A_0(2,1,3,4,5,6) & A(+--) &= A_0(2,1,3,4,5,6)^* \\
A(++-) &= A_0(1,2,3,4,6,5) & A(--+) &= A_0(1,2,3,4,6,5)^* \\
A(-+-) &= A_0(2,1,3,4,6,5) & A(+-+) &= A_0(2,1,3,4,6,5)^*.
\end{aligned} \qquad (8)$$

The numerator in the non-abelian diagrams can also be written in terms of the function $A$:

$$\begin{aligned}
&\bar{u}_\lambda(p_1)\gamma_\alpha u_\lambda(p_2)\ \bar{u}_\rho(p_3)\gamma_\mu u_\rho(p_4)\ \bar{u}_\sigma(p_5)\gamma_\nu u_\sigma(p_6) \\
&\times\ \{g^{\mu\alpha}(p_1+p_2)^\nu + g^{\alpha\nu}(p_5+p_6)^\mu + g^{\nu\mu}(p_3+p_4)^\alpha\} \\
&=\ A(\lambda,\rho,\sigma;1,2,3,4,5,6) - A(\sigma,\rho,\lambda;5,6,3,4,1,2)\ .
\end{aligned} \qquad (9)$$

Thus, for massless fermions, every helicity amplitude consists of a sum of very systematic, and relatively compact, expressions.

Extending now the $\gamma WW$ coupling with non-standard terms from the previous section we get as new numerator

$$[A(\lambda, \rho, \sigma; 1, 2, 3, 4, 5, 6) - A(\sigma, \rho, \lambda; 5, 6, 3, 4, 1, 2)]$$

$$+x_\gamma B(\lambda, \rho, \sigma; 1, 2, 3, 4, 5, 6)$$

$$+\frac{y_\gamma}{M_W^2}\left[(p_3 + p_4) \cdot (p_1 + p_2)B(\lambda, \sigma, \rho; 1, 2, 5, 6, 3, 4)\right.$$

$$-(p_5 + p_6) \cdot (p_1 + p_2)B(\lambda, \rho, \sigma; 1, 2, 3, 4, 5, 6)$$

$$\left.+(p_3 + p_4) \cdot (p_5 + p_6)B(\rho, \lambda, \sigma; 3, 4, 1, 2, 5, 6)\right]$$

$$+\frac{y_\gamma}{M_W^2}C(\lambda, \rho, \sigma; 1, 2, 3, 4, 5, 6)$$
(10)

and for the new $ZWW$-vertex

$$-(\cot\theta_w + \delta_Z)\left[A(\lambda, \rho, \sigma; 1, 2, 3, 4, 5, 6) - A(\sigma, \rho, \lambda; 5, 6, 3, 4, 1, 2)\right]$$

$$-x_Z B(\lambda, \rho, \sigma; 1, 2, 3, 4, 5, 6)$$

$$-\frac{y_Z}{M_W^2}\left[(p_3 + p_4) \cdot (p_1 + p_2)B(\lambda, \sigma, \rho; 1, 2, 5, 6, 3, 4)\right.$$

$$-(p_5 + p_6) \cdot (p_1 + p_2)B(\lambda, \rho, \sigma; 1, 2, 3, 4, 5, 6)$$

$$\left.+(p_3 + p_4) \cdot (p_5 + p_6)B(\rho, \lambda, \sigma; 3, 4, 1, 2, 5, 6)\right]$$

$$-\frac{y_Z}{M_W^2}C(\lambda, \rho, \sigma; 1, 2, 3, 4, 5, 6)$$

$$+\frac{iz_Z}{M_W^2}D(\lambda, \rho, \sigma; 1, 2, 3, 4, 5, 6)$$
(11)

where the new functions $B$, $C$ and $D$ are defined as

$$B(+,+,+; 1, 2, 3, 4, 5, 6) = 2\langle 31\rangle^*\langle 35\rangle^*\langle 26\rangle\langle 34\rangle$$
$$+2\langle 43\rangle^*\langle 15\rangle^*\langle 42\rangle\langle 46\rangle,$$
(12)

$$C(+,+,+; 1, 2, 3, 4, 5, 6) = 2\left[\langle 13\rangle^*\langle 23\rangle + \langle 14\rangle^*\langle 24\rangle\right] \times$$
$$\left[\langle 35\rangle^*\langle 45\rangle + \langle 36\rangle^*\langle 46\rangle\right]\left[\langle 51\rangle^*\langle 61\rangle + \langle 52\rangle^*\langle 62\rangle\right],$$
(13)

$$\begin{aligned}
D(+,+,+;1,2,3,4,5,6) = \\
& 2i\langle 35\rangle^*\langle 35\rangle^*\langle 34\rangle\langle 56\rangle \left[\langle 13\rangle^*\langle 23\rangle + \langle 14\rangle^*\langle 24\rangle\right] \\
& -2i\langle 46\rangle\langle 46\rangle\langle 34\rangle^*\langle 56\rangle^* \left[\langle 13\rangle^*\langle 23\rangle + \langle 14\rangle^*\langle 24\rangle\right] \\
& +2i\langle 31\rangle^*\langle 31\rangle^*\langle 34\rangle\langle 12\rangle \left[\langle 53\rangle^*\langle 63\rangle + \langle 54\rangle^*\langle 64\rangle\right] \\
& -2i\langle 42\rangle\langle 42\rangle\langle 34\rangle^*\langle 12\rangle^* \left[\langle 53\rangle^*\langle 63\rangle + \langle 54\rangle^*\langle 64\rangle\right].
\end{aligned} \qquad (14)$$

The expressions for $B$, $C$ and $D$ satisfy the same relations (8) as $A$. When equations (12)-(14) are denoted by $B_0$, $C_0$ and $D_0$ then

$$\begin{aligned}
B(+++) &= B_0(1,2,3,4,5,6) & B(---) &= B_0(1,2,3,4,5,6)^* \\
B(-++) &= B_0(2,1,3,4,5,6) & B(+--) &= B_0(2,1,3,4,5,6)^* \\
B(++-) &= B_0(1,2,3,4,6,5) & B(--+) &= B_0(1,2,3,4,6,5)^* \\
B(-+-) &= B_0(2,1,3,4,6,5) & B(+-+) &= B_0(2,1,3,4,6,5)^*
\end{aligned} \qquad (15)$$

and the same holds for $C$ and $D$.

Finally it should be noted that the vector boson propagators are implemented in the form $(q^2 - M_V^2 + iM_V\Gamma_V)^{-1}$, irrespective whether $q$ is timelike or not. This recipe guarantees the validity of electromagnetic gauge invariance. When this is violated even by a small amount forward electron cross sections can be off by orders of magnitude. This is due to photon exchange in the $t$-channel and was already noticed a long time ago as a problem in single $W$-production [19]. A less ad hoc solution to this problem is underway [20].

## 4 Results

Whereas at high energies total cross section measurements will give crucial information on the size of possible non-standard couplings, one has to consider at LEP2 angular distributions for this purpose. The natural five-dimensional differential cross section is

$$\frac{\mathrm{d}\sigma(e^-e^+ \longrightarrow W^-W^+ \longrightarrow f_1\bar{f}_2 f_3\bar{f}_4)}{\mathrm{d}\cos\theta\,\mathrm{d}\cos\theta_1\mathrm{d}\phi_1\mathrm{d}\cos\theta_2\mathrm{d}\phi_2} \qquad (16)$$

where $\theta$ is the angle between the incoming electron and $W^-$. The angles $\theta_1$, $\phi_1$ are the polar and azimuthal angles of the particle $f_1$ in the rest system of the parent particle $W^-$, whereas the angles $\theta_2$, $\phi_2$ fulfill a similar role for the antiparticle $\bar{f}_4$ originating from $W^+$. The angles are defined with respect to coordinate frames related to the $W^-$ and $W^+$. The z-directions are the directions of the momenta of the vectorbosons. The y-axes are defined by respectively $\vec{p}_- \times \vec{q}_-$ and $\vec{q}_+ \times \vec{p}_+$, where $\vec{p}_-, \vec{p}_+$ denote the momenta of the incoming electron and positron and $\vec{q}_-, \vec{q}_+$ the momenta of the $W^-$ and $W^+$.

In the zero width limit the above cross section is directly related to the helicity amplitudes for on-shell $W$-pair production and functions describing the decay of the vectorbosons [12, 13]. In principle direct fits to the above cross section could be performed. In practice one- or two-dimensional distributions will often be used. In the following we shall study $d\sigma/d\cos\theta$, $d\sigma/d\cos\theta_1$, $d\sigma/d\phi_1$.

The main purpose of this section is to illustrate effects of certain phenomena which have sofar not been incorporated in anomalous coupling studies. These are the effects of the finite $W$-width, of ISR and of background diagrams. It is useful to define a number of (differential) cross sections $\sigma$ evaluated under different assumptions. In the first place we introduce SM cross sections $\sigma_{SM,on}$, $\sigma_{SM}$, $\sigma_{SM,ISR}$ and $\sigma_{SM,all}$ which are respectively on-shell, off-shell signal cross sections (i.e. with three diagrams), the off-shell signal case with ISR and the cross section containing all diagrams. Furthermore, we define $\sigma_{AN}$, $\sigma_{AN,ISR}$, and $\sigma_{AN,all}$ which are (differential) anomalous cross sections calculated with the three signal diagrams, without or with ISR, and with all diagrams without ISR.

The following ratios give an illustration of the effects of the finite width, the ISR, background diagrams and of non-standard couplings:

$$R_1 = \frac{\sigma_{SM}}{\sigma_{SM,on}}, \tag{17}$$

$$R_2 = \frac{\sigma_{SM,ISR}}{\sigma_{SM}}, \tag{18}$$

$$R_3 = \frac{\sigma_{SM,all}}{\sigma_{SM}}, \tag{19}$$

$$R_4 = \frac{\sigma_{AN}}{\sigma_{SM}}, \tag{20}$$

$$R_5 = \frac{\sigma_{AN,all}}{\sigma_{SM,all}}. \tag{21}$$

The reaction which we take as example is

$$e^- e^+ \longrightarrow e^- \bar{\nu}_e u \bar{d}. \tag{22}$$

The following input parameters are used

$$\alpha^{-1} = 128.07$$
$$\sin^2\theta_w = 0.23103$$

$$\begin{aligned} M_Z &= 91.1888 \text{ GeV} \\ M_W &= 80.23 \text{ GeV} \\ \Gamma_Z &= 2.4974 \text{ GeV} \\ \Gamma_W &= 2.08 \text{ GeV}. \end{aligned}$$

where $\sin^2 \theta_w$ is chosen in such a way that combined with the above running $\alpha$ value the correct $G_\mu$ is obtained:

$$G_\mu = \frac{\alpha \pi}{\sqrt{2} \sin^2 \theta_w M_W^2}. \tag{23}$$

For the ISR the usual value of $\alpha$ is used. It should be noted that the above experimental values for the total widths are incorporated in the propagators. In EXCALIBUR the decay widths of the $W$ into a lepton pair or quark pair are independent from the input total width. They follow from the other input parameters. Since one would like to have $s$-dependent widths in the $s$-channel and because this would violate gauge invariance the following practical procedure is used. The s-dependent widths can be transformed into a constant width [21]. When this constant width is used in both $s$- and $t$-channel gauge invariance is ensured in a simple way, which numerically agrees well with theoretically more sound methods [20]. Thus the calculations are performed with propagators $(q^2 - \tilde{M}_V^2 + i\tilde{M}_V \tilde{\Gamma}_V)^{-1}$, where

$$\begin{aligned} \tilde{M}_V &= M_V/\sqrt{1 + \gamma_V^2}, & (24) \\ \tilde{\Gamma}_V &= \Gamma_V/\sqrt{1 + \gamma_V^2}, & (25) \\ \gamma_V &= \Gamma_V/M_V. & (26) \end{aligned}$$

With these input values various differential cross sections have been evaluated. The SM and anomalous cross sections with all diagrams have to be calculated with cuts avoiding thus the singularities due to the massless fermions. In order to make meaningful comparisons the cross section $\sigma_{SM}$ in $R_3$ has the same cuts. The imposed cuts are

$$\begin{aligned} E_{e^-, u, \bar{d}} &> 20 \text{ GeV} & (27) \\ |\cos \theta_{e^-, u, \bar{d}}| &< 0.9 & (28) \\ |\cos \angle(u\bar{d})| &< 0.9 & (29) \\ m_{u\bar{d}} &> 10 \text{ GeV} & (30) \end{aligned}$$

where $\theta$ is the angle between the outgoing particles and the incoming beams.

In tables 4–7 total cross sections $\sigma_{AN}$, $\sigma_{AN,ISR}$, $\sigma_{AN}(\text{cuts})$, $\sigma_{AN,all}$ are listed for an energy of 190 GeV. The SM differential cross sections are given in figure 2. Differential cross section ratios are given in the form of histograms in figs 3–10. From $R_1$ it is seen that the inclusion of the finite width already changes the $\cos\theta$ distribution by a few percent. Comparing the on- and off-shell $\sigma_{AN}$ a similar angular modification arises [22]. Similarly the inclusion of ISR or background diagrams introduce even larger modifications of this angular distribution. In order to show the effects of the various anomalous couplings histograms of $R_4$ and $R_5$ are presented with values $\pm 0.5$ for every coupling successively, the others being zero at the same time. When doing the analysis with the three signal diagrams both for SM and non-standard couplings ($R_4$) the overall picture is roughly the same as for the case where both cross sections contain all diagrams ($R_5$). The effects of the anomalous couplings show up most clearly in the $\cos\theta$ distribution as can been seen when comparing to the pictures of the $\cos\theta_1$ and $\phi_1$ distributions.

## 5 Conclusions

With the extended EXCALIBUR program it becomes possible to study effects of anomalous couplings in all four-fermion final states which receive contributions from non-abelian diagrams. In this way finite width effects of the vectorbosons are incorporated and studies of ISR and background diagrams can be made. Up to 2 TeV the program works efficiently. For studies at higher energies the present phase space treatment of the multiperipheral massive vectorboson diagrams should be adjusted, which in principle does not pose any problem. For LEP2 this is not yet required.

From the presented results it is clear that in particular the distribution in the $W$ production angle $\theta$ is affected by the finite $W$ width, ISR and background. Also here anomalous couplings show up most clearly. The results of this paper give a quantative assesment of the above effects, which have hitherto not been considered in the literature.

## Acknowledgements

We are grateful to W. Beenakker, M. Bilenky, R. Kleiss, J.L. Kneur and G.J. van Oldenborgh for valuable dicussions.

| label | final state | $N_a$ | $N_n$ | total | signals |
|---|---|---|---|---|---|
| 1 | $e^+e^-\nu_e\bar{\nu}_e$ | 48 | 8 | 56 | 1 2 3 |
| 2 | $e^-\bar{\nu}_e\nu_\mu\mu^+$ | 14 | 4 | 18 | 1 2 |
| 3 | $e^-\bar{\nu}_e\nu_\tau\tau^+$ | | | | |
| 4 | $\nu_e e^+\mu^-\bar{\nu}_\mu$ | | | | |
| 5 | $\nu_e e^+\tau^-\bar{\nu}_\tau$ | | | | |
| 6 | $\mu^+\mu^-\nu_\mu\bar{\nu}_\mu$ | 17 | 2 | 19 | 1 |
| 7 | $\tau^+\tau^-\nu_\tau\bar{\nu}_\tau$ | | | | |
| 8 | $\mu^-\bar{\nu}_\mu\nu_\tau\tau^+$ | 7 | 2 | 9 | 1 |
| 9 | $\tau^-\bar{\nu}_\tau\nu_\mu\mu^+$ | | | | |
| 10 | $\nu_e\bar{\nu}_e\mu^+\mu^-$ | 17 | 2 | 19 | 3 |
| 11 | $\nu_e\bar{\nu}_e\tau^+\tau^-$ | | | | |
| 12 | $\nu_e\bar{\nu}_e\nu_e\bar{\nu}_e$ | 32 | 4 | 36 | 3 |
| 13 | $\nu_e\bar{\nu}_e\nu_\mu\bar{\nu}_\mu$ | 11 | 1 | 12 | 3 |
| 14 | $\nu_e\bar{\nu}_e\nu_\tau\bar{\nu}_\tau$ | | | | |

Table 1: leptonic four-fermion final states in $e^+e^-$ collisions.

| label | final state | $N_a$ | $N_n$ | total | signals |
|---|---|---|---|---|---|
| 1 | $e^-\bar{\nu}_e u\bar{d}$ | 16 | 4 | 20 | 1 2 |
| 2 | $e^-\bar{\nu}_e c\bar{s}$ | | | | |
| 3 | $\nu_e e^+ d\bar{u}$ | | | | |
| 4 | $\nu_e e^+ s\bar{c}$ | | | | |
| 5 | $\mu^-\bar{\nu}_\mu u\bar{d}$ | 8 | 2 | 10 | 1 |
| 6 | $\mu^-\bar{\nu}_\mu c\bar{s}$ | | | | |
| 7 | $\mu^+\nu_\mu d\bar{u}$ | | | | |
| 8 | $\mu^+\nu_\mu s\bar{c}$ | | | | |
| 9 | $\tau^-\bar{\nu}_\tau u\bar{d}$ | | | | |
| 10 | $\tau^-\bar{\nu}_\tau c\bar{s}$ | | | | |
| 11 | $\tau^+\nu_\tau d\bar{u}$ | | | | |
| 12 | $\tau^+\nu_\tau s\bar{c}$ | | | | |
| 13 | $\nu_e\bar{\nu}_e u\bar{u}$ | 17 | 2 | 19 | 3 |
| 14 | $\nu_e\bar{\nu}_e c\bar{c}$ | | | | |
| 15 | $\nu_e\bar{\nu}_e d\bar{d}$ | 17 | 2 | 19 | 3 |
| 16 | $\nu_e\bar{\nu}_e s\bar{s}$ | | | | |
| 17 | $\nu_e\bar{\nu}_e b\bar{b}$ | | | | |

Table 2: semileptonic four-fermion final states in $e^+e^-$ collisions

| label | final state | $N_a$ | $N_n$ | total | signals |
|---|---|---|---|---|---|
| 1 | $u\bar{u}d\bar{d}$ | 33 | 2 | 35 | 1 |
| 2 | $c\bar{c}d\bar{d}$ | | | | |
| 3 | $u\bar{d}s\bar{c}$ | 9 | 2 | 11 | 1 |
| 4 | $d\bar{u}c\bar{s}$ | | | | |

Table 3: hadronic four-fermion final states in $e^+e^-$ collisions.

| $\sqrt{s} = 190\ GeV$: $\sigma_{SM} = 0.6490 \pm 0.0010$ pb |||||||
|---|---|---|---|---|---|---|
| only $WW$-diagrams, no ISR |||||||
|  | $\delta_Z$ | $x_\gamma$ | $y_\gamma$ | $x_Z$ | $y_Z$ | $z_Z$ |
| $-0.5$ | 0.6729 | 0.6866 | 0.6823 | 0.6582 | 0.6592 | 0.6863 |
|  | 0.0011 | 0.0015 | 0.0011 | 0.0011 | 0.0011 | 0.0011 |
| $-0.2$ | 0.6537 | 0.6611 | 0.6578 | 0.6512 | 0.6501 | 0.6573 |
|  | 0.0011 | 0.0010 | 0.0010 | 0.0010 | 0.0010 | 0.0011 |
| $0.2$ | 0.6504 | 0.6402 | 0.6456 | 0.6486 | 0.6509 | 0.6493 |
|  | 0.0010 | 0.0010 | 0.0011 | 0.0010 | 0.0010 | 0.0010 |
| $0.5$ | 0.6666 | 0.6358 | 0.6524 | 0.6534 | 0.6617 | 0.6679 |
|  | 0.0010 | 0.0010 | 0.0010 | 0.0010 | 0.0010 | 0.0010 |

Table 4: $\sigma_{AN}$ succesively calculated for all anomalous couplings vanishing but one. Each second row gives the error to $\sigma_{AN}$.

| $\sqrt{s} = 190\ GeV$: $\sigma_{SM} = 0.5790 \pm 0.0011$ pb |||||||
|---|---|---|---|---|---|---|
| only $WW$-diagrams, ISR |||||||
|  | $\delta_Z$ | $x_\gamma$ | $y_\gamma$ | $x_Z$ | $y_Z$ | $z_Z$ |
| $-0.5$ | 0.6001 | 0.6084 | 0.6053 | 0.5882 | 0.5878 | 0.6077 |
|  | 0.0011 | 0.0011 | 0.0011 | 0.0011 | 0.0011 | 0.0011 |
| $-0.2$ | 0.5825 | 0.5889 | 0.5861 | 0.5806 | 0.5797 | 0.5853 |
|  | 0.0011 | 0.0011 | 0.0011 | 0.0011 | 0.0011 | 0.0011 |
| $0.2$ | 0.5809 | 0.5721 | 0.5766 | 0.5792 | 0.5811 | 0.5794 |
|  | 0.0010 | 0.0011 | 0.0011 | 0.0011 | 0.0011 | 0.0010 |
| $0.5$ | 0.5945 | 0.5683 | 0.5827 | 0.5834 | 0.5900 | 0.5928 |
|  | 0.0011 | 0.0011 | 0.0011 | 0.0011 | 0.0011 | 0.0011 |

Table 5: $\sigma_{AN,ISR}$ for similar case as in table 4

| $\sqrt{s} = 190\ GeV$: $\sigma_{SM} = 0.45970 \pm 0.00097$ pb | | | | | | |
|---|---|---|---|---|---|---|
| only $WW$-diagrams, no ISR, cuts | | | | | | |
| | $\delta_Z$ | $x_\gamma$ | $y_\gamma$ | $x_Z$ | $y_Z$ | $z_Z$ |
| $-0.5$ | 0.4783 | 0.48705 | 0.48484 | 0.46781 | 0.46721 | 0.4909 |
| | 0.0010 | 0.00099 | 0.00099 | 0.00098 | 0.00098 | 0.0010 |
| $-0.2$ | 0.46372 | 0.46863 | 0.46623 | 0.46191 | 0.46067 | 0.46787 |
| | 0.00098 | 0.00098 | 0.00097 | 0.00097 | 0.00097 | 0.00099 |
| $0.2$ | 0.46050 | 0.45361 | 0.45774 | 0.45930 | 0.46134 | 0.45898 |
| | 0.00097 | 0.00097 | 0.00097 | 0.00097 | 0.00097 | 0.00097 |
| $0.5$ | 0.47080 | 0.44998 | 0.46330 | 0.46256 | 0.46921 | 0.47068 |
| | 0.00097 | 0.00097 | 0.00097 | 0.00097 | 0.00097 | 0.00098 |

Table 6: $\sigma_{AN}$ with the imposed cuts. Cases as in table 4

| $\sqrt{s} = 190\ GeV$: $\sigma_{SM} = 0.4705 \pm 0.0010$ pb | | | | | | |
|---|---|---|---|---|---|---|
| all diagrams, no ISR, cuts | | | | | | |
| | $\delta_Z$ | $x_\gamma$ | $y_\gamma$ | $x_Z$ | $y_Z$ | $z_Z$ |
| $-0.5$ | 0.4881 | 0.4961 | 0.4942 | 0.4791 | 0.4778 | 0.5018 |
| | 0.0011 | 0.0010 | 0.0010 | 0.0010 | 0.0010 | 0.0011 |
| $-0.2$ | 0.4743 | 0.4786 | 0.4767 | 0.4724 | 0.4708 | 0.4782 |
| | 0.0010 | 0.0010 | 0.0010 | 0.0010 | 0.0010 | 0.0010 |
| $0.2$ | 0.4724 | 0.4665 | 0.4690 | 0.4701 | 0.4725 | 0.4699 |
| | 0.0010 | 0.0010 | 0.0010 | 0.0010 | 0.0010 | 0.0010 |
| $0.5$ | 0.4824 | 0.4651 | 0.4758 | 0.4737 | 0.4808 | 0.4816 |
| | 0.0010 | 0.0010 | 0.0010 | 0.0010 | 0.0010 | 0.0010 |

Table 7: $\sigma_{AN,all}$ with the imposed cuts. Cases as in table 4

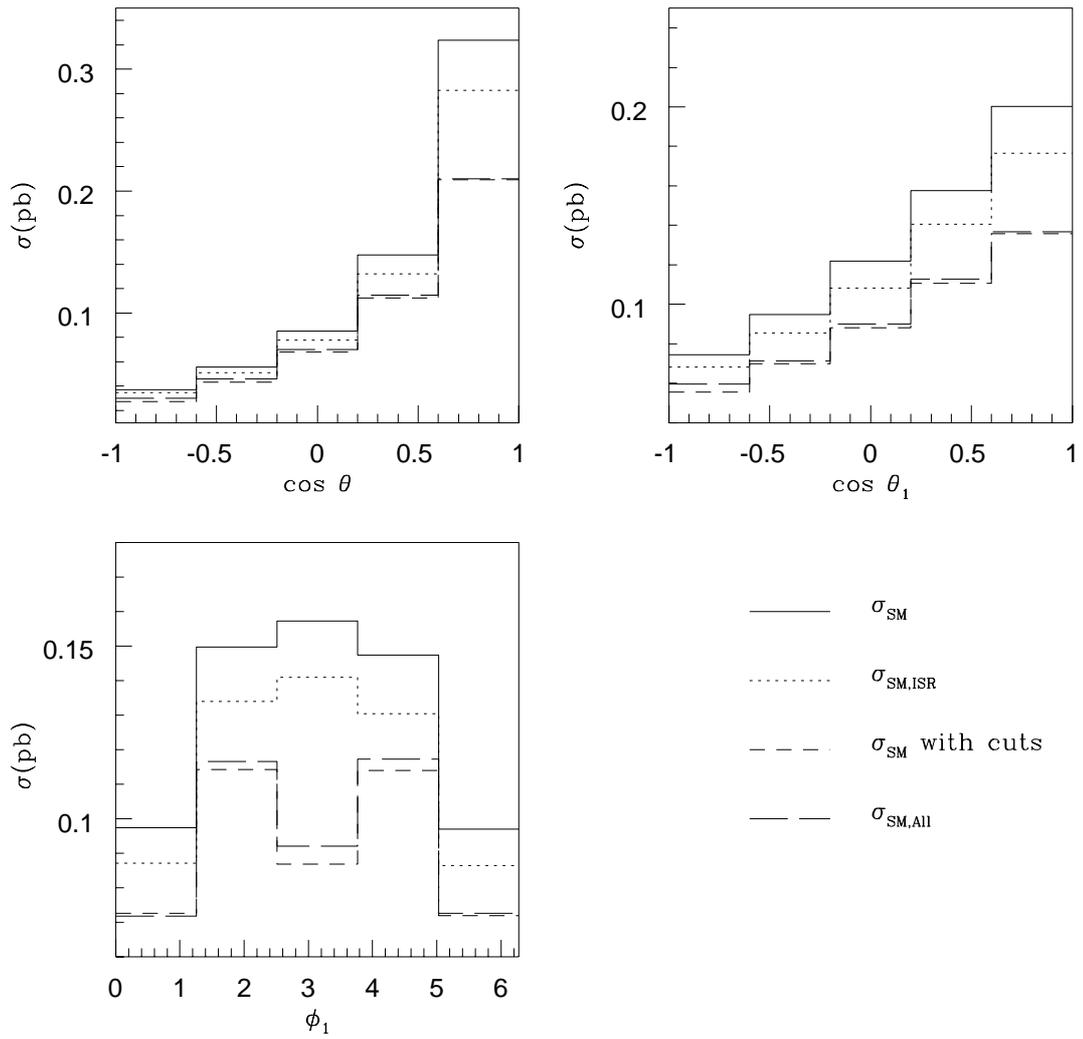

Figure 2: Distributions for the Standard Model

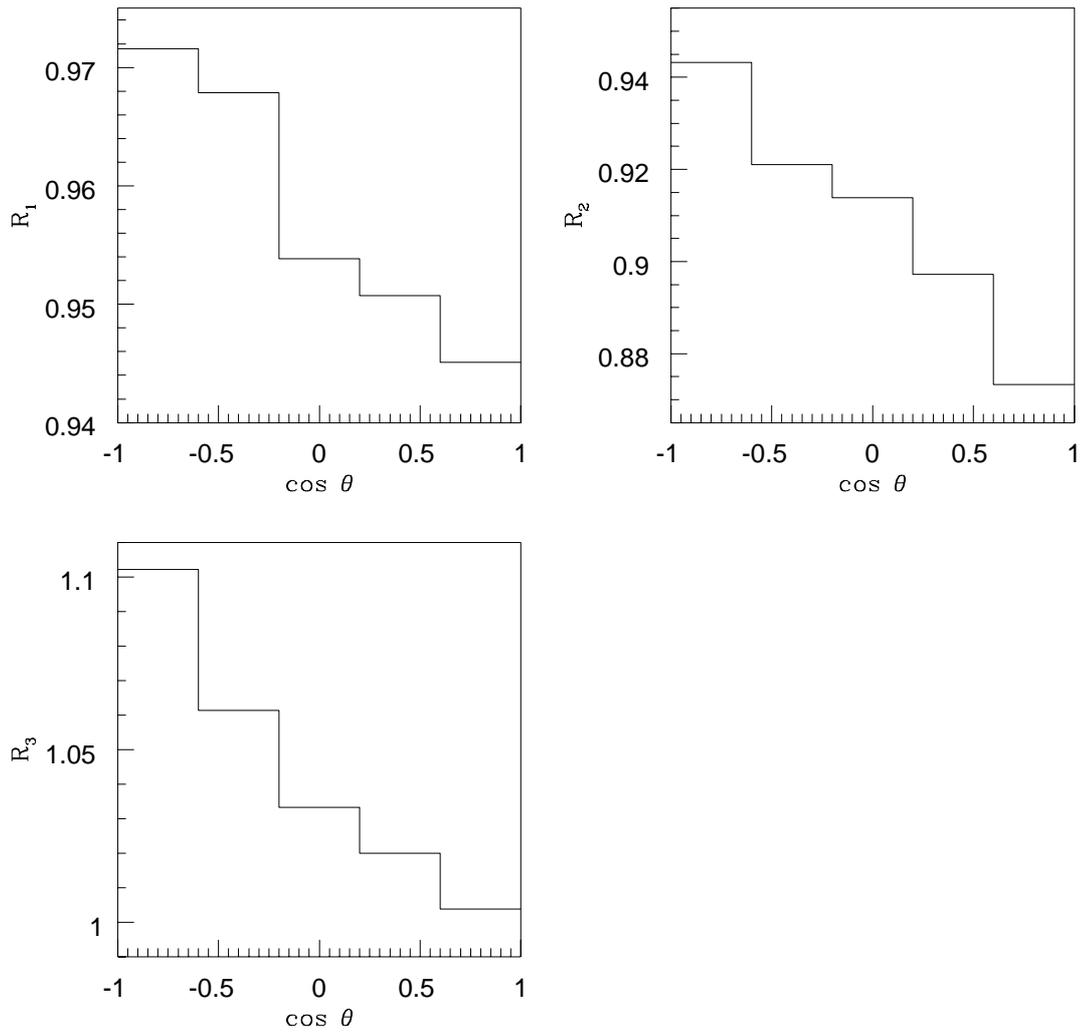

Figure 3: The ratios $R_1$,$R_2$,$R_3$

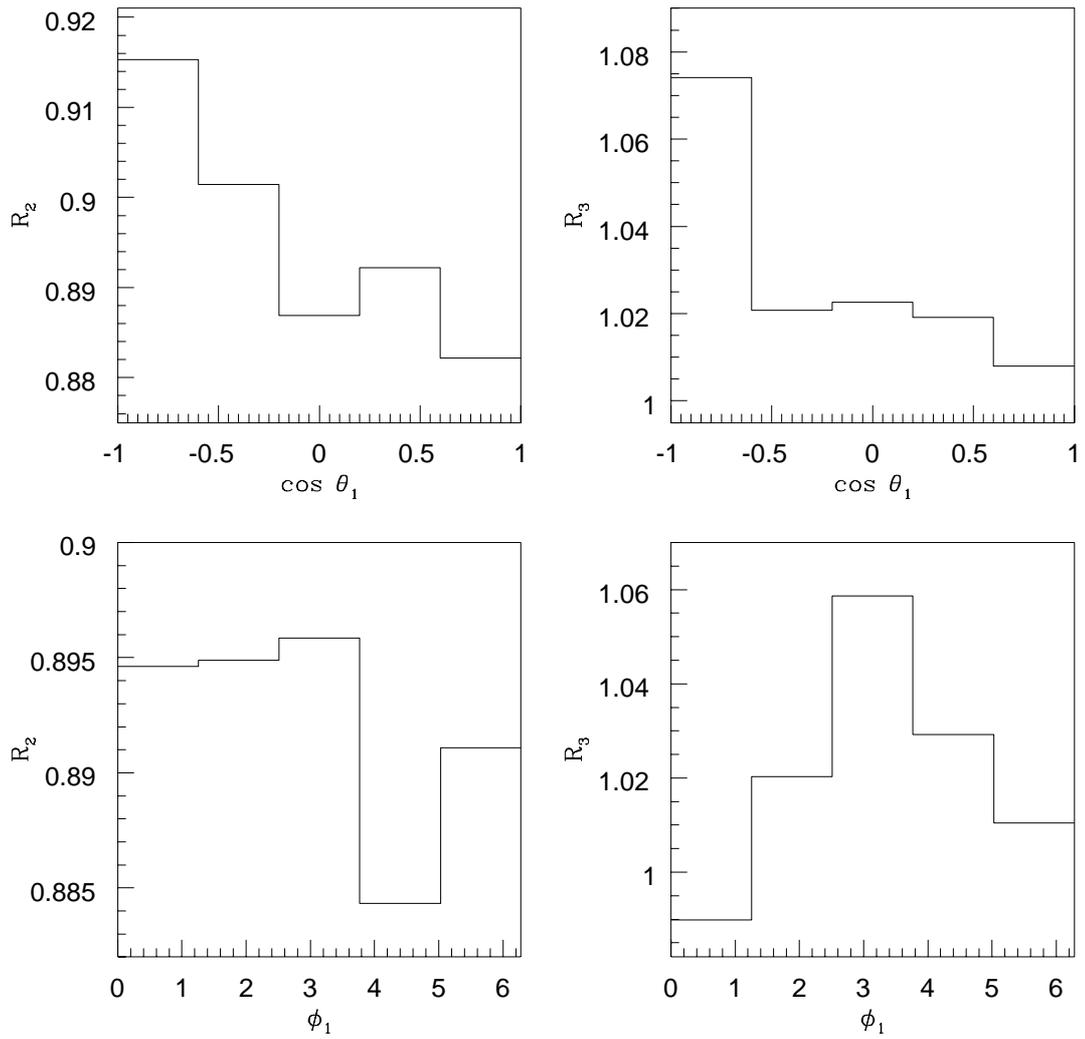

Figure 4: The ratios $R_2, R_3$

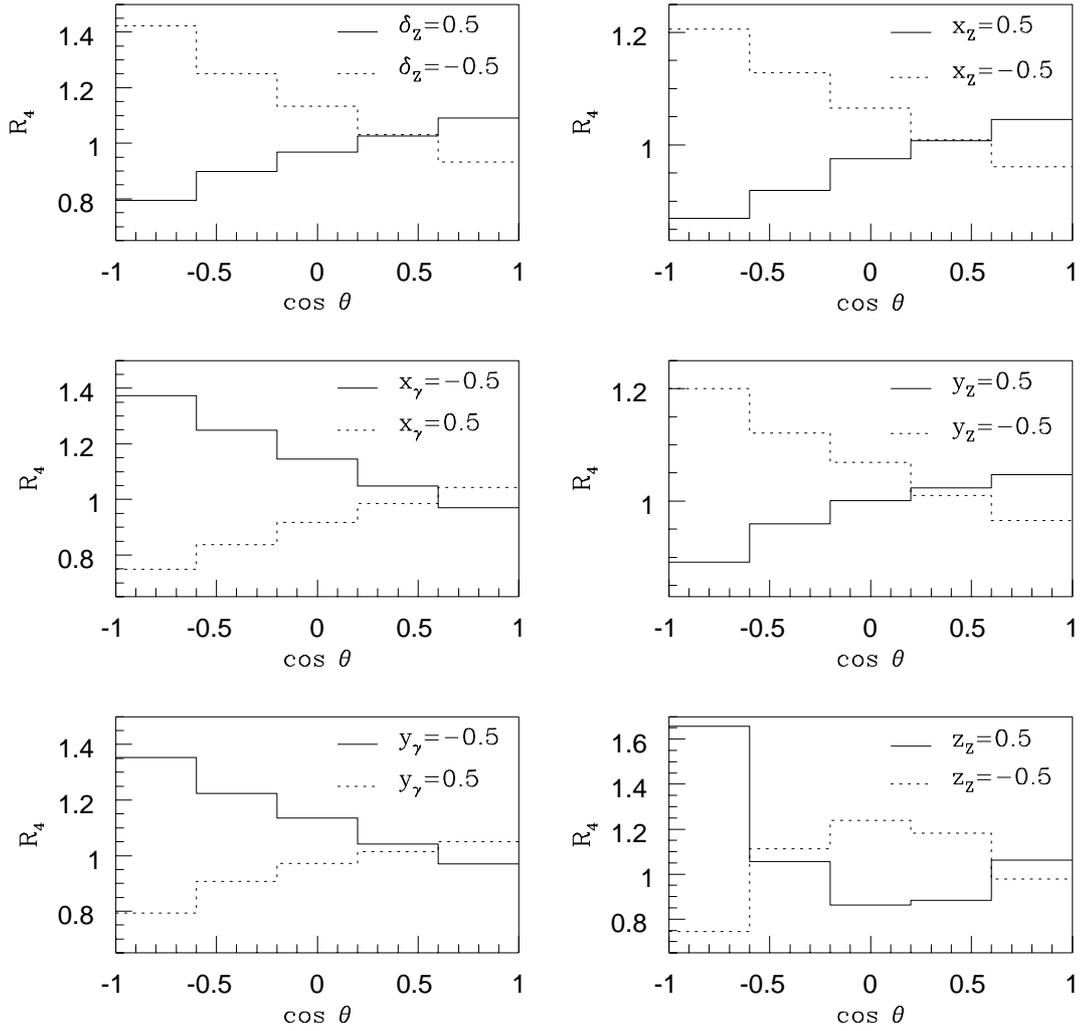

Figure 5: The ratio $R_4$

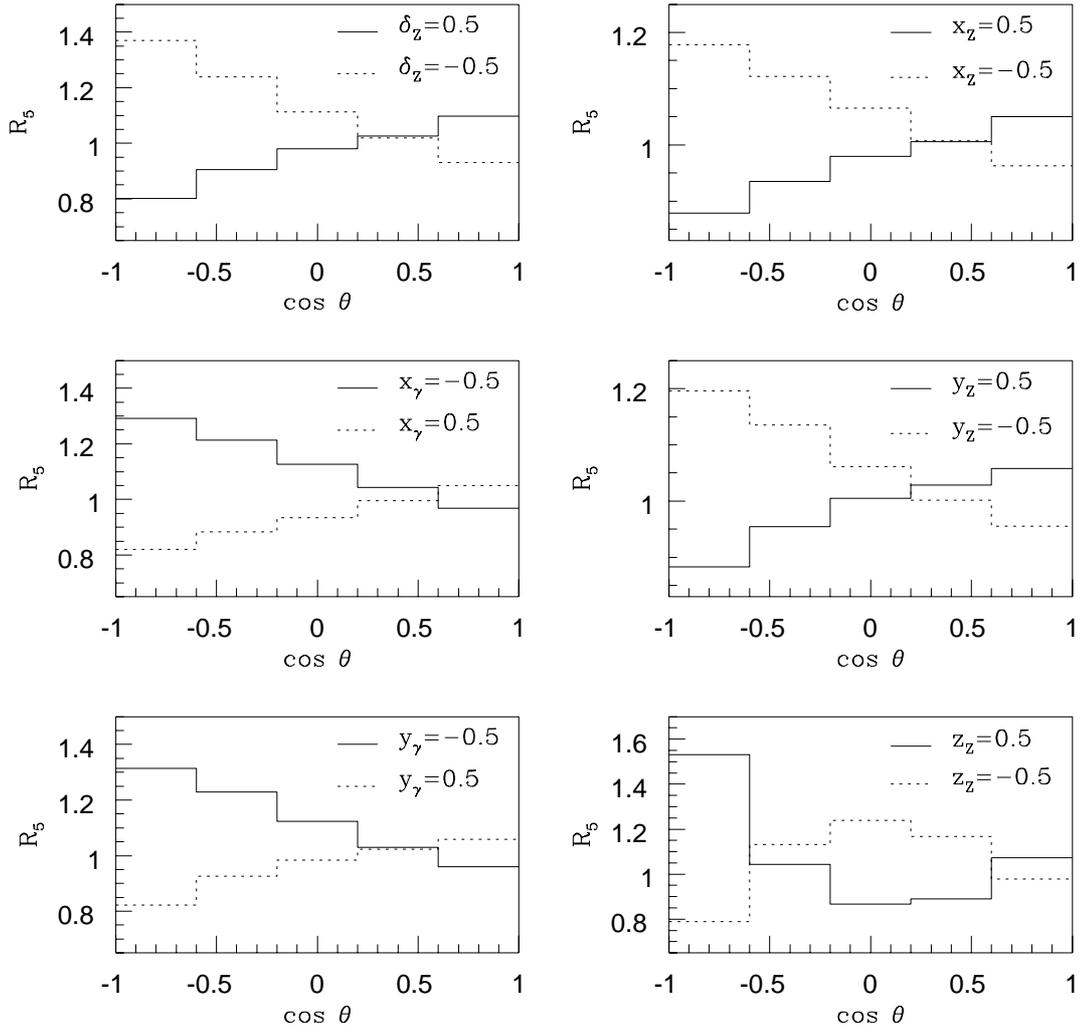

Figure 6: The ratio $R_5$

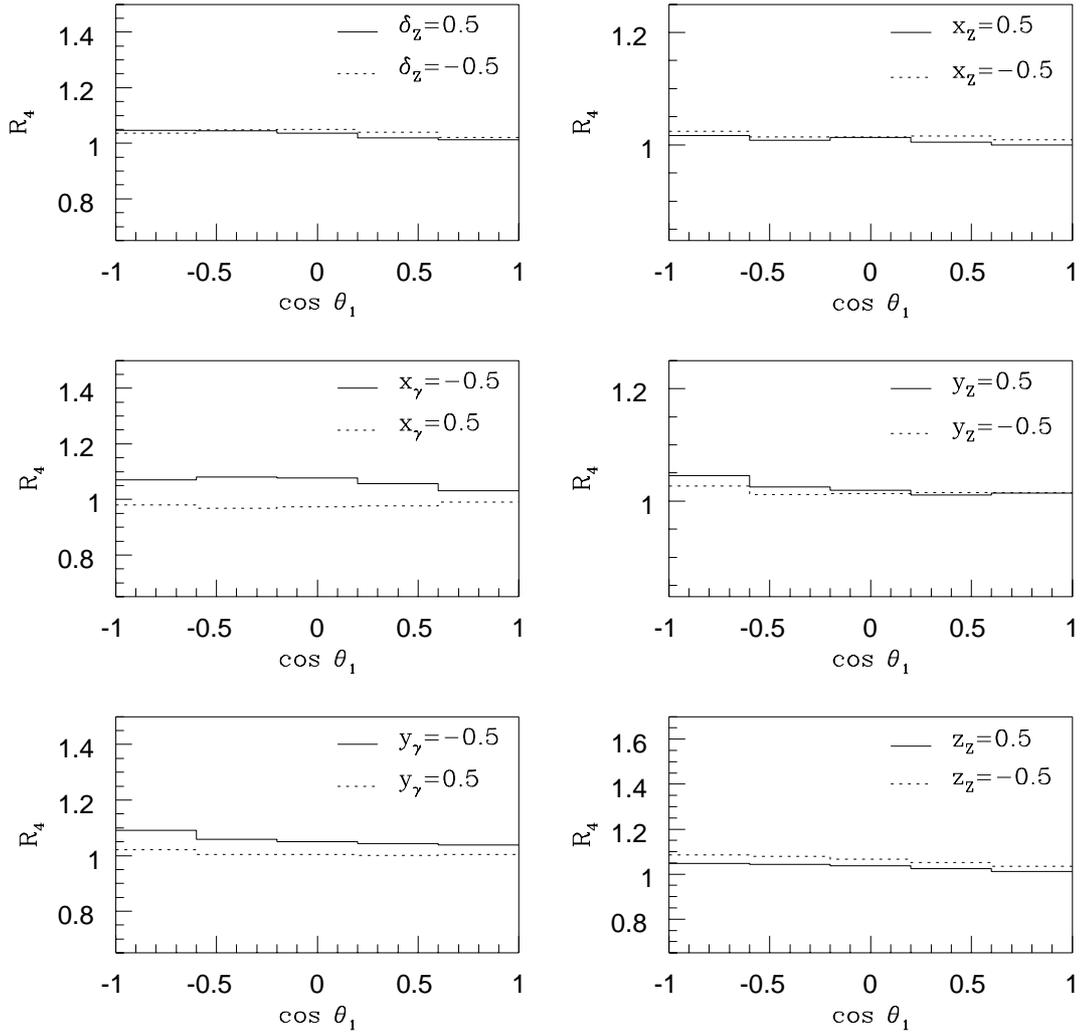

Figure 7: The ratio $R_4$

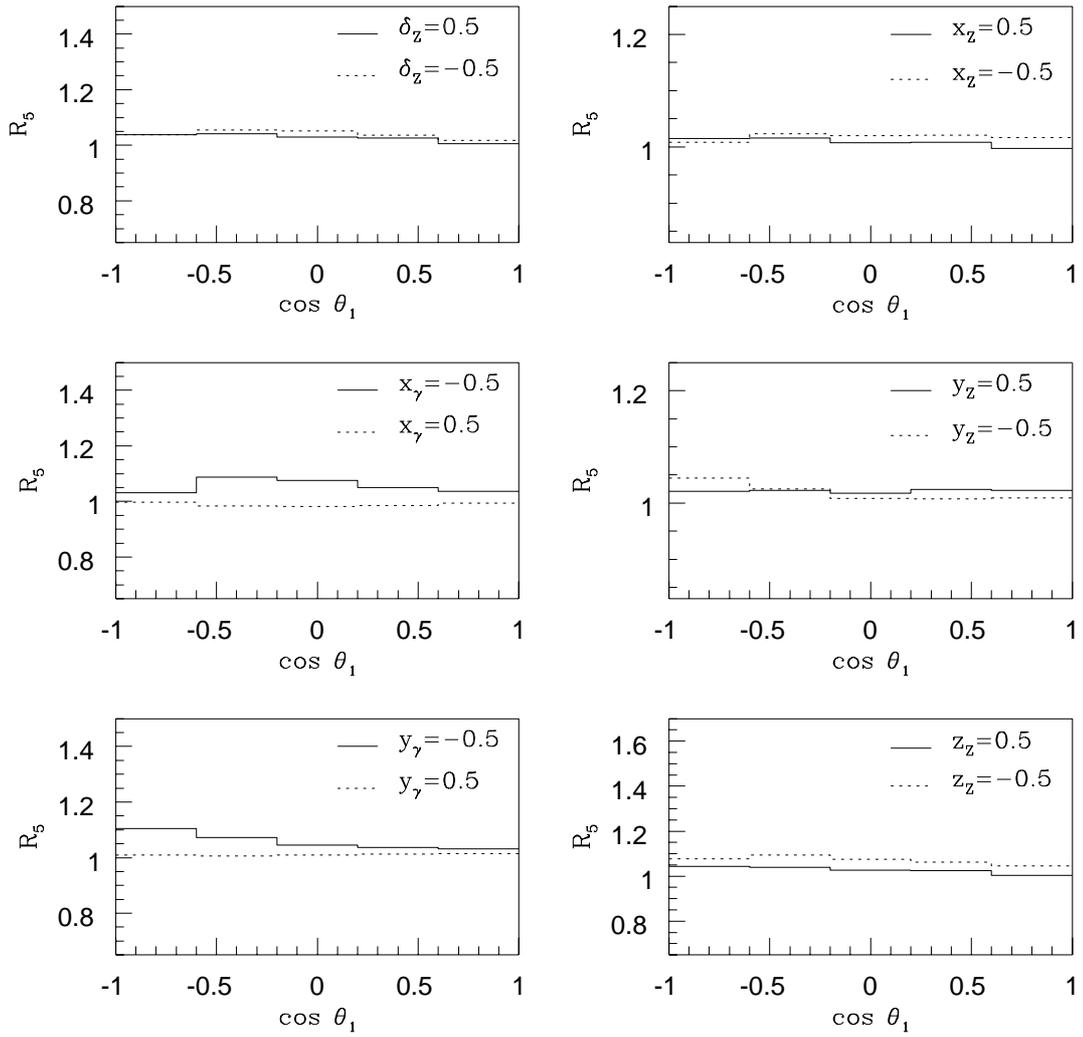

Figure 8: The ratio $R_5$

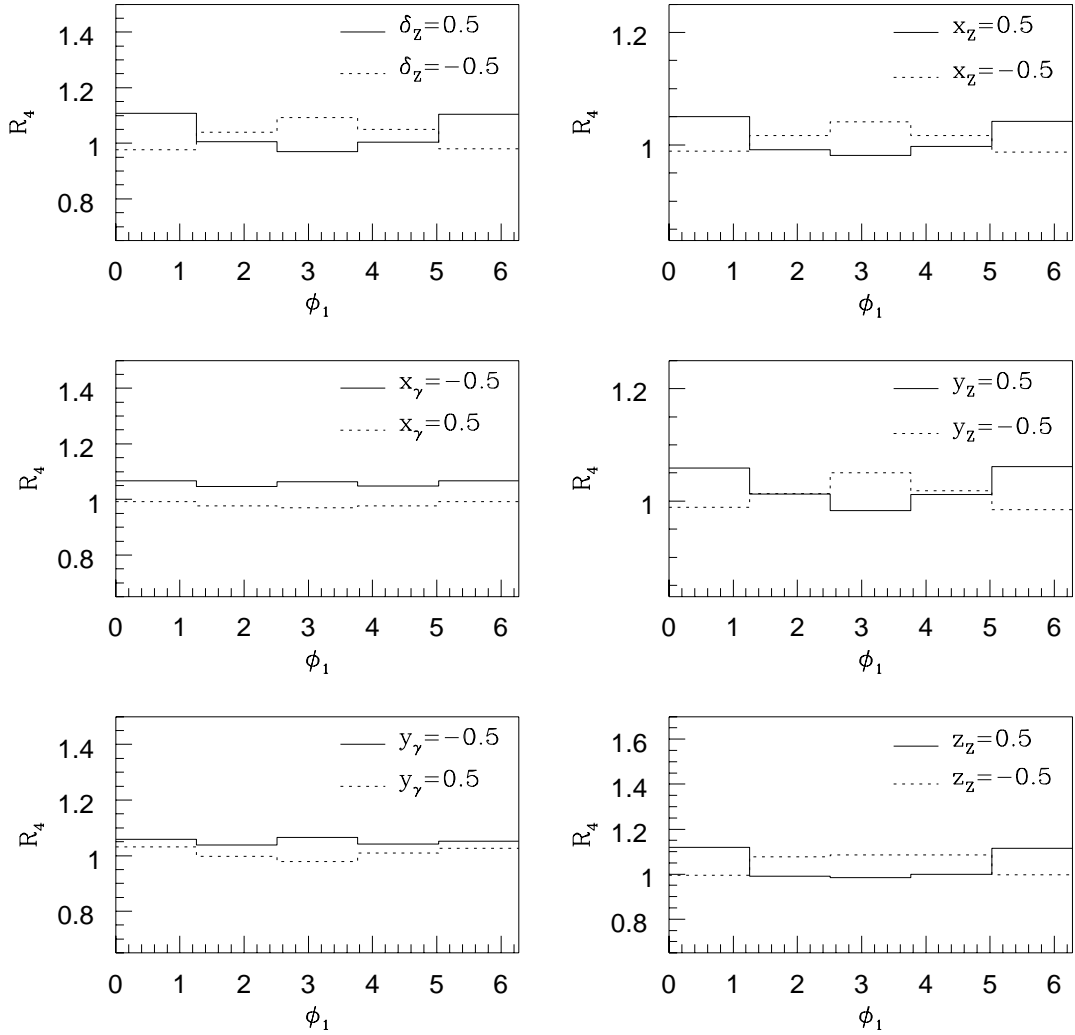

Figure 9: The ratio $R_4$

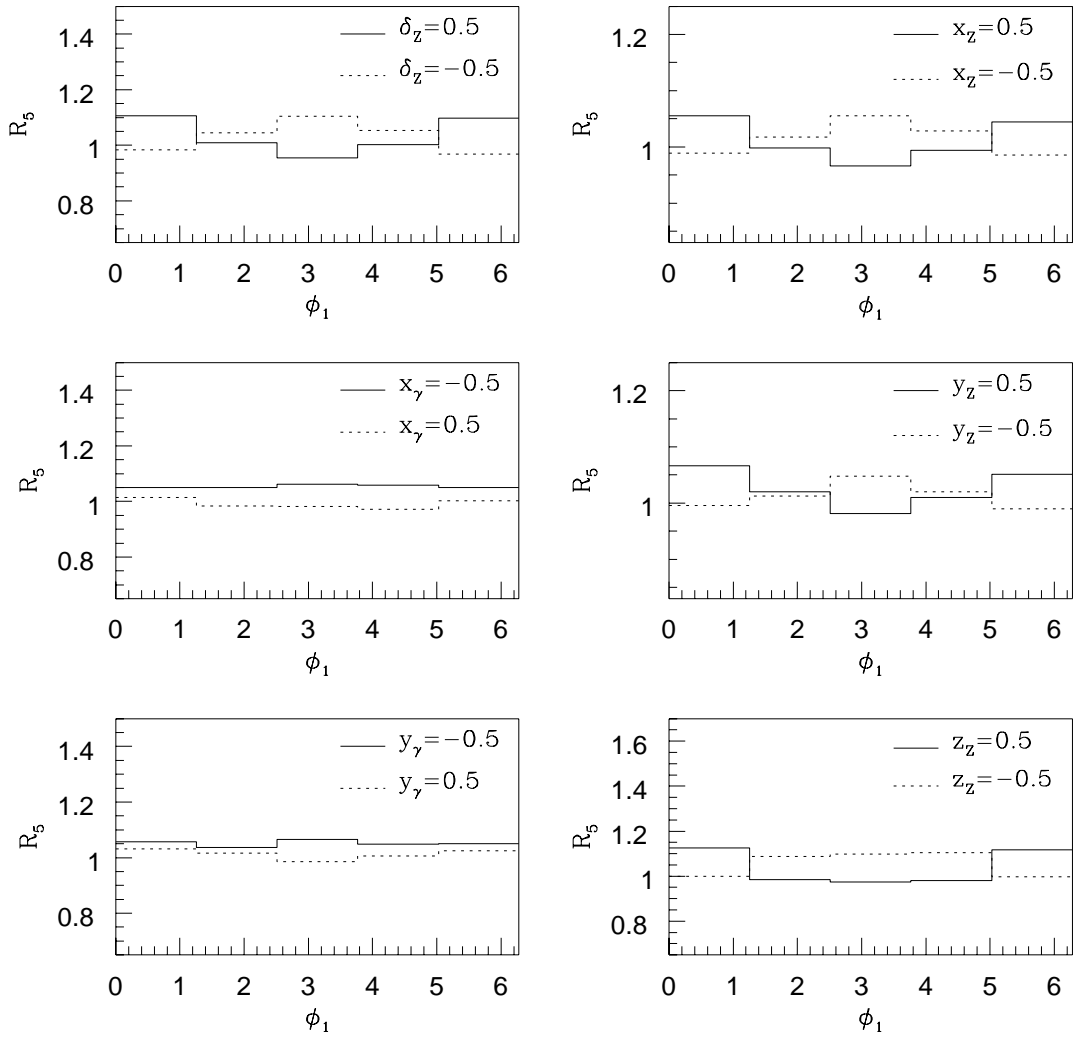

Figure 10: The ratio $R_5$